\documentclass[a4paper]{article}

\usepackage{INTERSPEECH2022}
\usepackage{lipsum}
\usepackage{booktabs}
\usepackage{threeparttable}
\usepackage{caption}
\usepackage{subcaption}
\usepackage{multirow}

\title{Separating Content from Speaker Identity in Speech for the Assessment of Cognitive Impairments}
\name{Dongseok Heo$^{1,2}$, Cheul Young Park$^1$, Jaemin Cheun$^1$, Myung Jin Ko$^1$}
\address{
  $^1$Silvia Health Inc., Seoul, Republic of Korea\\
  $^2$Department of Computer Science and Engineering, Seoul National University
}
\email{\{dongseok,cheul,jaemin,myungjin\}@silviahealth.com}

\begin{document}

\maketitle
\begin{abstract}
Deep speaker embeddings have been shown effective for assessing cognitive impairments aside from their original purpose of speaker verification. However, the research found that speaker embeddings encode speaker identity and an array of information, including speaker demographics, such as sex and age, and speech contents to an extent, which are known confounders in the assessment of cognitive impairments. In this paper, we hypothesize that content information separated from speaker identity using a framework for voice conversion is more effective for assessing cognitive impairments and train simple classifiers for the comparative analysis on the DementiaBank Pitt Corpus. Our results show that while content embeddings have an advantage over speaker embeddings for the defined problem, further experiments show their effectiveness depends on information encoded in speaker embeddings due to the inherent design of the architecture used for extracting contents.
\end{abstract}

\section{Introduction}

Language degradation is prevalent in Alzheimer’s dementia (AD), which amounts to 60 to 80\% of dementia cases, starting from semantics and later progressing to phonetic and syntactic aspects~\cite{kirshner2012primary}. Not confined to AD, language impairment is observed in many cognitive diseases such as cerebrovascular dementia~\cite{vuorinen2000common} and Parkinson’s~\cite{lewis1998language}.

Speech with rich information across multiple modalities including semantics, linguistics, and acoustics enables comprehensive examination~\cite{fraser2016linguistic}. Computational analysis of speech has advantages over manual analysis as computational models can incorporate domain experts’ knowledge to provide cost-efficient but exact assessments. Researchers are active in applying this data-driven approach for the assessment of cognitive impairments, as demonstrated in recent community events such as the ADReSS(o) challenge~\cite{Luz2020,luz21_interspeech}.

One approach for the automated assessment of cognitive decline with speech is to use speaker identity vectors, or speaker embeddings, which are mainly used for speaker verification. Speaker embeddings have been shown to contain diverse information in addition to the speaker identity, including the content of spoken language, gender, speaking rate, word order, and utterance length~\cite{wang2017does,ghahremani2018end,raj2019probing}. From that, we can expect speaker embeddings to encode information related to cognitive abilities and be used to assess cognitive impairments. Researchers indeed have shown that models trained with the speaker embeddings can distinguish patients from healthy individuals and predict the severity of impairment via Mini-Mental Status Evaluation (MMSE) scores~\cite{pappagari2020using,pappagari2021automatic,jeancolas2021x}.

However, no known work has investigated the validity of speaker embeddings for assessing cognitive decline to the best of our knowledge. As per their intended usage, speaker embeddings encode information characterizing each individual, for example acoustic qualities of speech such as timbre, pitch, speaking rate, and demographics like age, gender, and ethnicity. These individual characteristics nevertheless can confound the assessment of cognitive impairments~\cite{anderson2004ethnic,podcasy2016considering}, while research suggests that correcting confounding variables like age can result in a more accurate assessment~\cite{dukart2011age}.

Inspired by this, we explore whether normalizing individual-specific characteristics and focusing on the content aspect of speech can result in a more accurate assessment of cognitive impairments via speech. In this work, we examine the effectiveness of content embeddings for assessing cognitive impairments with the DementiaBank dataset~\cite{becker1994natural}. Using an autoencoder-based framework for voice conversion, we separate speech content from speaker identity and extract as embeddings to use them for the binary classification task between individuals with various symptoms of cognitive impairment and healthy control group.

\begin{table}[t]
\caption{DementiaBank Pitt Corpus statistics, w/ and w/o aug.}
\label{tab:dementiabank}
\centering
\renewcommand{\arraystretch}{1.2}
\begin{tabular}{@{}rcccc@{}}
\toprule
\multicolumn{1}{l}{} & \multicolumn{2}{c}{\textbf{Control (CN)}} & \multicolumn{2}{c}{\textbf{Impaired (IM)}} \\ \midrule
\textit{(no aug.)} & \textbf{Mean $\pm$ SD} & \textbf{Range} & \textbf{Mean $\pm$ SD} & \textbf{Range} \\ \midrule
Age & $64.0\pm8.1$ & {[}46, 80{]} & $71.2\pm8.6$ & {[}49, 88{]} \\
MMSE & $29.1\pm1.1$ & {[}26, 30{]} & $19.9\pm5.5$ & {[}3, 30{]} \\
M:F & \multicolumn{2}{c}{39:59 (N=98)} & \multicolumn{2}{c}{68:126 (N=194)} \\ \midrule
\textit{(aug.)} & \multicolumn{1}{l}{\textbf{Mean $\pm$ SD}} & \multicolumn{1}{l}{\textbf{Range}} & \multicolumn{1}{l}{\textbf{Mean $\pm$ SD}} & \multicolumn{1}{l}{\textbf{Range}} \\ \midrule
Age & $64.7\pm7.7$ & {[}46, 81{]} & $71.4\pm8.4$ & {[}49, 90{]} \\
MMSE & $29.1\pm1.1$ & {[}24, 30{]} & $19.7\pm5.6$ & {[}1, 30{]} \\
M:F & \multicolumn{2}{c}{87:155 (N=242)} & \multicolumn{2}{c}{120:190 (N=310)} \\ \bottomrule
\end{tabular}
\end{table}

\section{The DementiaBank Pitt Corpus}

\begin{figure*}[t]
  \centering
  \begin{subfigure}[t]{0.59\linewidth}
    \centering
    \includegraphics[width=\linewidth]{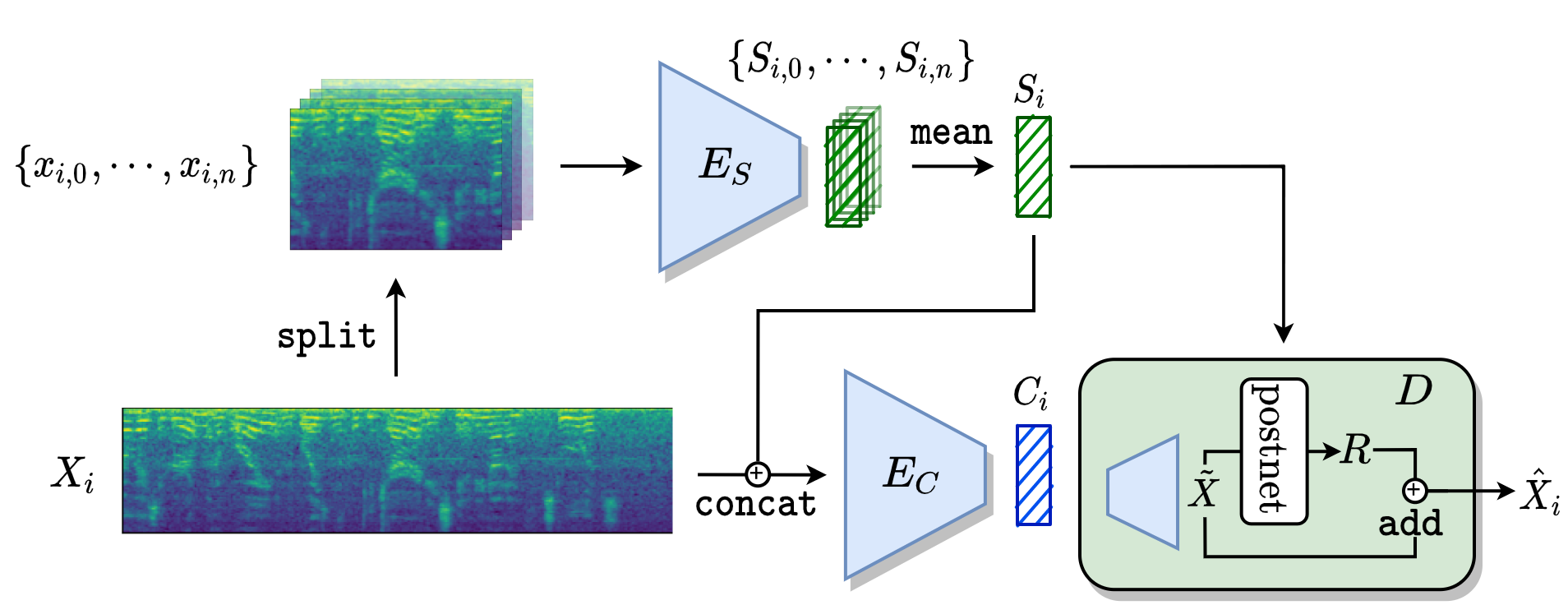}
    \caption{Overall training scheme for content separation}
    \label{subfig:encoder-decoder}
  \end{subfigure}
  \hfill
  \begin{subfigure}[t]{0.175\linewidth}
    \centering
    \includegraphics[width=\linewidth]{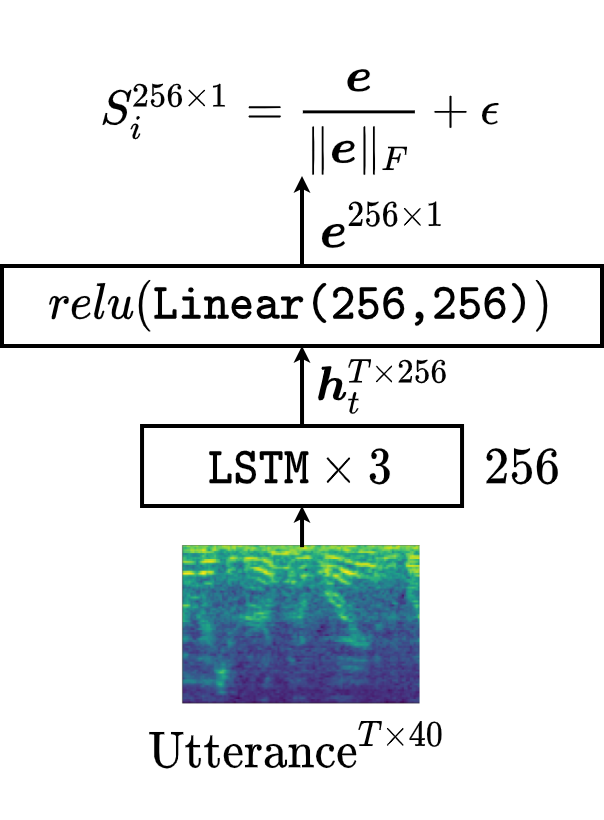}
    \caption{Speaker encoder $E_S$}
    \label{subfig:speaker-encoder}
  \end{subfigure}
  \hfill
  \begin{subfigure}[t]{0.225\linewidth}
    \centering
    \includegraphics[width=\linewidth]{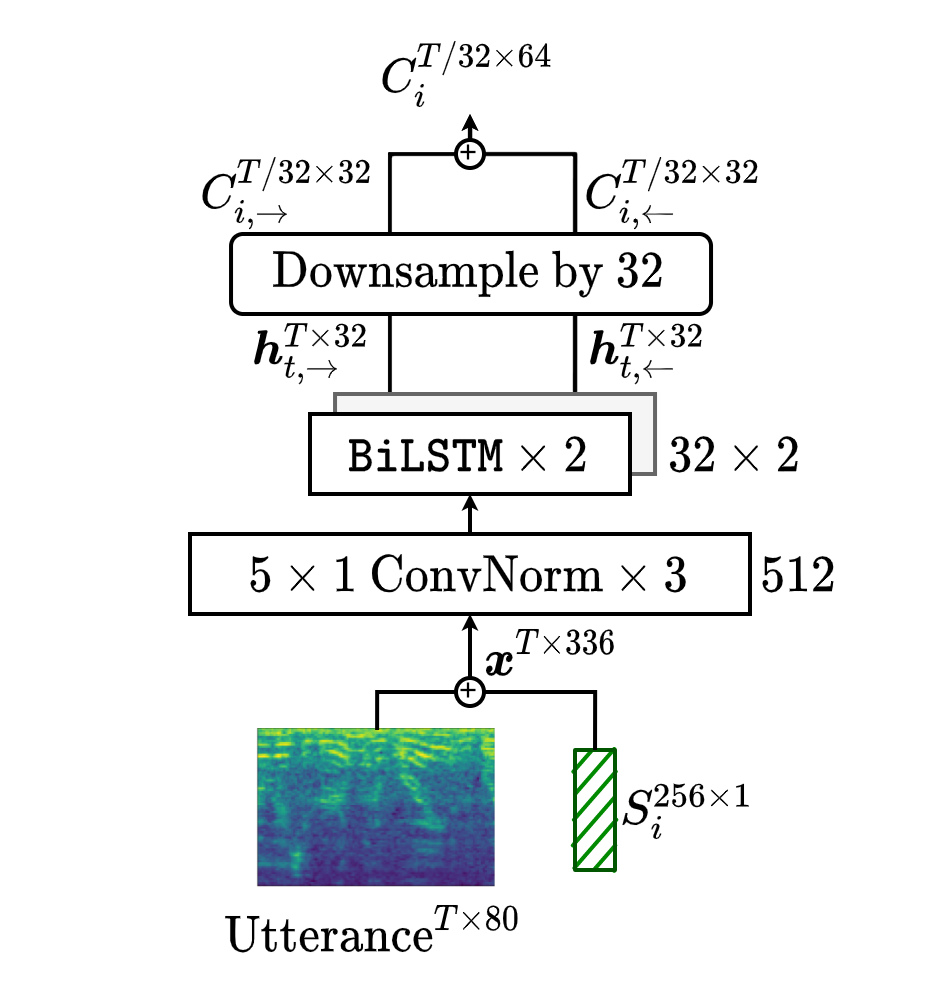}
    \caption{Content encoder $E_C$}
    \label{subfig:content-encoder}
  \end{subfigure}
  \caption{Encoder-decoder architecture for content embedding separation based on the AutoVC framework.}
  \label{fig:framework}
\end{figure*}

DementiaBank Pitt Corpus (Table~\ref{tab:dementiabank}) is the largest publicly available dataset of cognitively impaired speeches. Speech samples were collected longitudinally on a yearly basis between 1983 and 1988 from healthy individuals (control group, CN) and individuals with varying symptoms and severity of cognitive impairment (impaired group, IM), responding to four speech-based tasks. In this paper, we use speech responses to the picture description task using the Boston Cookie Theft picture~\cite{goodglass2001bdae} from 98 controls and 194 patients diagnosed with cognitive impairment. Content embeddings are extracted from speech samples, and their effectiveness for assessing cognitive impairment is compared with speaker embeddings. Further, we augment the dataset by considering speeches collected from a single individual but at different sessions as unique samples, thus increasing the total size of the dataset from 292 to 552, with 242 controls and 310 impaired. This is to let classifiers learn better by increasing the number of trainable examples, given the relatively small size of the corpus.

\section{Content separation framework}

We use the AutoVC framework~\cite{qian2019autovc}, a many-to-many non-parallel voice conversion framework based on the autoencoder architecture, to separate contents from speaker characteristics and extract them as embeddings from speech. The framework comprises three modules, content encoder, pretrained speaker encoder, and decoder. See Figure~\ref{fig:framework} for the overall architecture of the content embedding extraction framework with the respective details of speaker and content encoders.

Where $X_i$ is a mel spectrogram of size $N \times T$ computed from a speech of speaker $i$, with $N$ as the number of mel frequency bins (which we set to 80, with a window of length 64ms and step size 16ms) and $T$ as the number of time steps, a content encoder $E_C$ takes $X_i$ concatenated with a speaker embedding $S_i$ at each time step (frame) as an input and returns a content embedding $C_i$.

Similarly, a speaker encoder $E_S$ returns a speaker identity embedding $S_i$ given a speech of speaker $i$ in the form of a 2-dimensional mel spectrogram as an input. However, speaker embeddings we use for training in our implementation are averages of utterance-level speaker embeddings; i.e., for each speaker, we compute a set of speaker embeddings where each embedding corresponds to an utterance comprising a full speech of a speaker and take their arithmetic mean to obtain a unique speaker embedding. The intuition behind this is to normalize the content information encoded in speaker embeddings by taking their average to let the content encoder better separate speech content from speaker identity, as speaker identity should remain constant while contents differ across utterances from the same speaker.

Thus, assuming a scenario of extracting a content embedding $C_i$ for speaker $i$ who produces speech $X_i$, which is comprised of a set of partial utterances $\{x_{i,0}, \cdots, x_{i,n}\}$, the process is formulated as follows:
\begin{equation}
  C_i = E_C(X_i, S_i)
  \label{eq1}
\end{equation}
where
\begin{equation}
  S_i = E_S(X_i) = \frac{1}{n}(E_S(x_{i,0}) + \cdots + E_S(x_{i,n}))
  \label{eq2}
\end{equation}
Outcomes from the content encoder are two $32 \times T/32$ vectors $C_{i,\rightarrow}$ and $C_{i,\leftarrow}$, which are each forward and backward pass outputs of a 2-layer bidirectional LSTM, downsampled in time domain $T$ by a factor of 32. Combining them results in a $64 \times T/32$ vector $C_i$.

However, training the content encoder requires a decoder $D$ that reconstructs speech by combining speaker embedding $S_i$ and content embedding $C_i$ to estimate a source mel spectrogram $\hat{X}_{i\rightarrow i}$. The reconstruction takes two steps; first, an initial approximation of a source spectrogram $\tilde{X}_{i \rightarrow i}$, followed by an estimation of a residual signal $R$ with a post-network as suggested in~\cite{shen2018natural}. The final reconstruction result is the addition of an initial estimate and a residual. Thus, the formulation for speech reconstruction with the decoder is as follows:
\begin{equation}
    \hat{X}_{i \rightarrow i} = \tilde{X}_{i \rightarrow i} + R = D(C_i, S_i)
    \label{eq3}
\end{equation}
Note that while a decoder can be used to transfer the style/identity of another speaker $j$ to the speech content provided by a source speaker $i$, ex) $\hat{X}_{i \rightarrow j} = D(C_i, S_j)$, we only focus on training with speaker and content embeddings of the same speaker as the rest is irrelevant for extracting content embeddings.

Finally, only the encoder-decoder part is optimized during training while the speaker encoder is assumed to be pretrained and fixed, with a loss function defined as the following:
\begin{equation}
    \min_{E_C(\cdot, \cdot), D(\cdot, \cdot)} L = L_\text{recon} + \mu \cdot L_\text{recon0} + \lambda \cdot L_\text{cont}
    \label{eq4}
\end{equation}
where
\begin{equation}
    L_\text{recon} = \texttt{mean}(|{\hat{X}_{i \rightarrow i} - X_i}|)
    \label{eq5}
\end{equation}
is an l1 loss with a mean reduction between a post-network processed reconstructed speech and a source speech,
\begin{equation}
    L_\text{recon0} = \texttt{mean}(|\tilde{X}_{i \rightarrow i} - X_i|)
    \label{eq6}
\end{equation}
is similarly an l1 loss with a mean reduction between an initial approximation of a reconstructed speech and a source speech,
\begin{equation}
    L_\text{cont} = |E_C(\hat{X}_{i \rightarrow i}, S_i) - C_i|
    \label{eq7}
\end{equation}
and eqn.~\ref{eq7} is an l1 loss between a reconstructed content embedding and a source content embedding, but without reduction.

While the initial reconstruction loss and the content reconstruction loss can be parameterized with $\mu$ and $\lambda$, we set them to 1 in our implementation for simplicity.

\section{Training speaker encoder}
For a speaker encoder $E_S$, we use the same architecture and preprocessing steps as the original AutoVC paper, which was initially suggested in the generalized end-to-end (GE2E) loss for training speaker verification models~\cite{wan2018generalized}.

Audio files are first resampled to 16,000Hz and loudness normalized to -30dBFS. After silent segments are smoothed out with Voice Activity Detection (VAD), speeches are then transformed to mel spectrograms of window length 25ms, step size 10ms, and 40 log-mel-filterbank channels. Finally, with 10 random utterances chosen for each speaker, partial frames of length 1,600ms (160 frames) are extracted at random from utterances and stacked as an array of shape $[\text{num\_speakers} \times \text{num\_utterances}, \text{num\_frames}, \text{dim\_mel\_channels}]$.

\begin{figure}[t]
  \centering
  \begin{subfigure}[b]{0.49\linewidth}
    \centering
    \includegraphics[width=\linewidth]{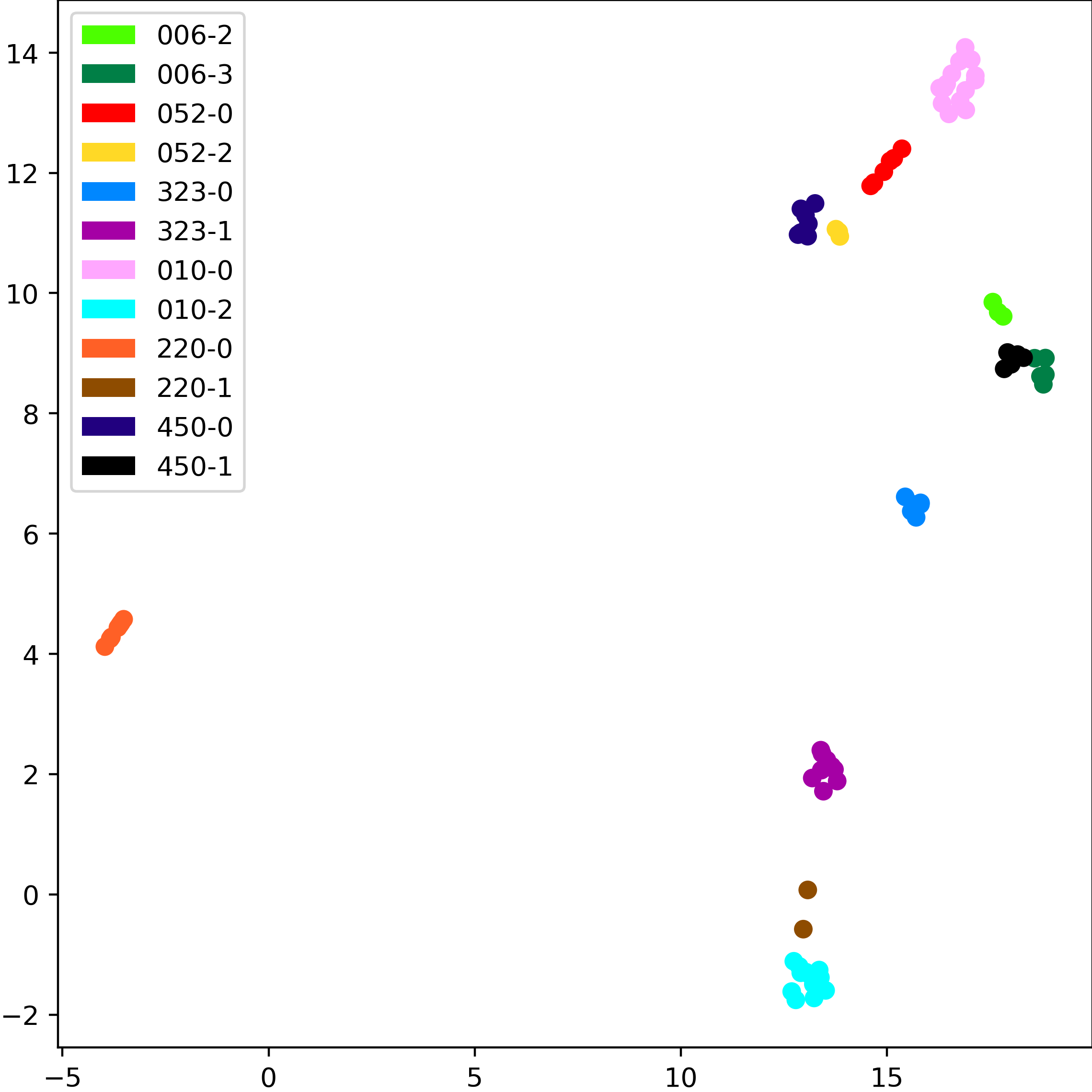}
    \caption{Fully trained}
    \label{subfig:fully-trained}
  \end{subfigure}
  \hfill
  \begin{subfigure}[b]{0.49\linewidth}
    \centering
    \includegraphics[width=\linewidth]{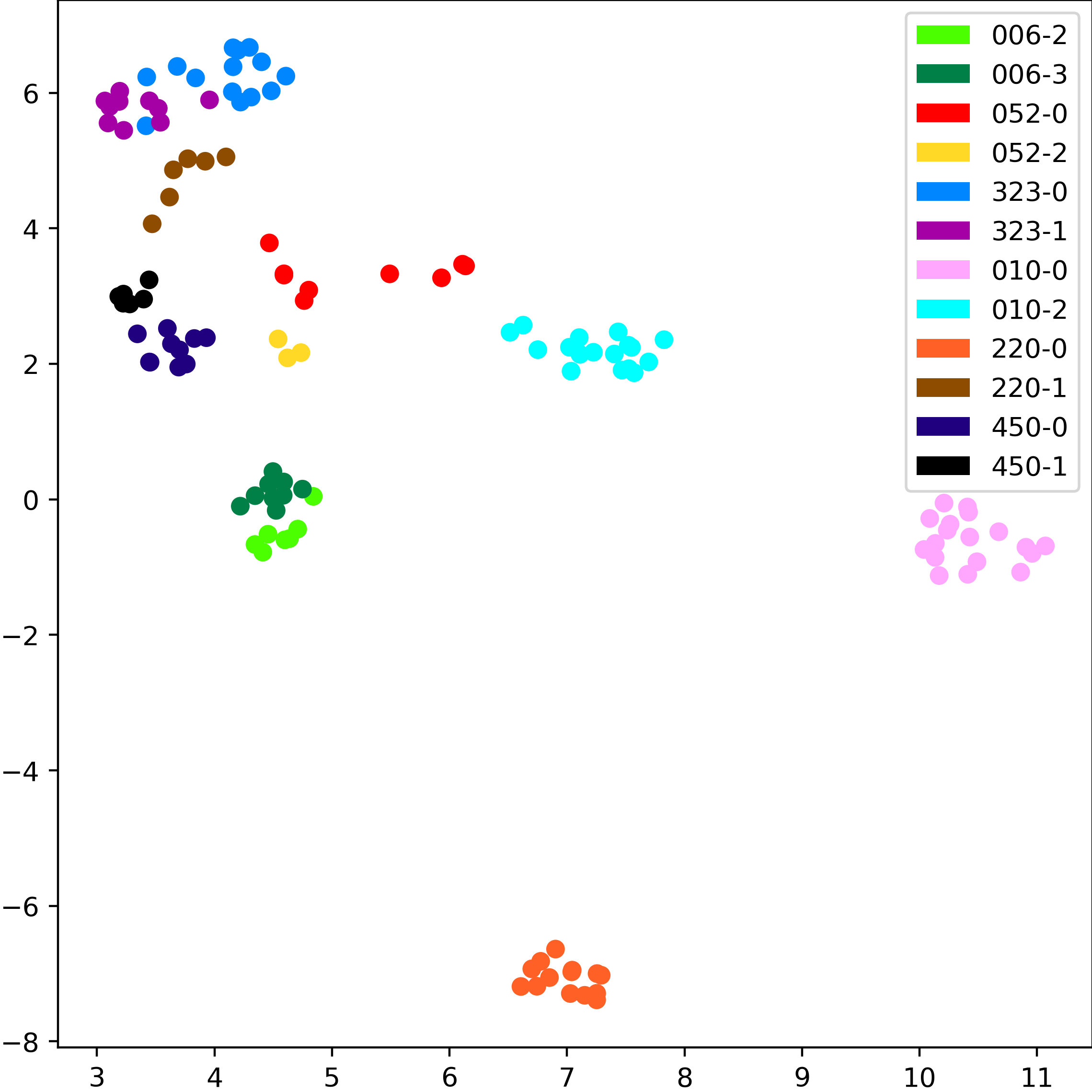}
    \caption{Finetuned}
    \label{subfig:finetuned}
  \end{subfigure}
  \caption{UMAP-based clustering of speaker embeddings from utterances of 6 randomly chosen speakers: CN=$[006, 052, 323]$ and IM=$[010, 220, 450]$}
  \label{fig:clutering}
\end{figure}

Our implementation uses a speaker encoder pretrained on a publicly available speech corpora, comprised of the training set from LibriSpeech~\cite{panayotov2015librispeech}, VoxCeleb1 Dev A - D~\cite{nagrani17_interspeech}, and VoxCeleb2~\cite{chung18b_interspeech}, resulting in 3,201 hours of data with 8,371 different speakers. We finetune the final linear output layer and similarity scaling parameters of the pretrained speaker encoder with the DementiaBank Pitt corpus. We empirically observed that fully training a model end-to-end without freezing parameters results in a very low equal error rate (EER) below 0.005, but a sufficiently low EER of ~0.03 can also be achieved with finetuning. Visualizing utterance-level speaker embeddings via the Uniform Manifold Approximation and Projection (UMAP)~\cite{mcinnes2018umap} as in Figure~\ref{fig:clutering} also shows finetuning results in better speaker embeddings clusters at an utterance level than a full training.

\section{Experimental setup}
We compare content embeddings against speaker embeddings for assessing cognitive impairments using the cookie theft speech recordings from the DementiaBank Pitt Corpus. Evaluation is set up as a binary classification task between a cognitively impaired group (IM) and a healthy control group (CN). We hypothesize that content embeddings are more effective for assessing cognitive impairments than speaker embeddings as individual characteristics confounding the identification of impairments are normalized in the process contents are separated from speaker characteristics in speech.

\begin{figure}[t]
  \centering
  \begin{subfigure}[b]{0.32\linewidth}
    \centering
    \includegraphics[width=\linewidth]{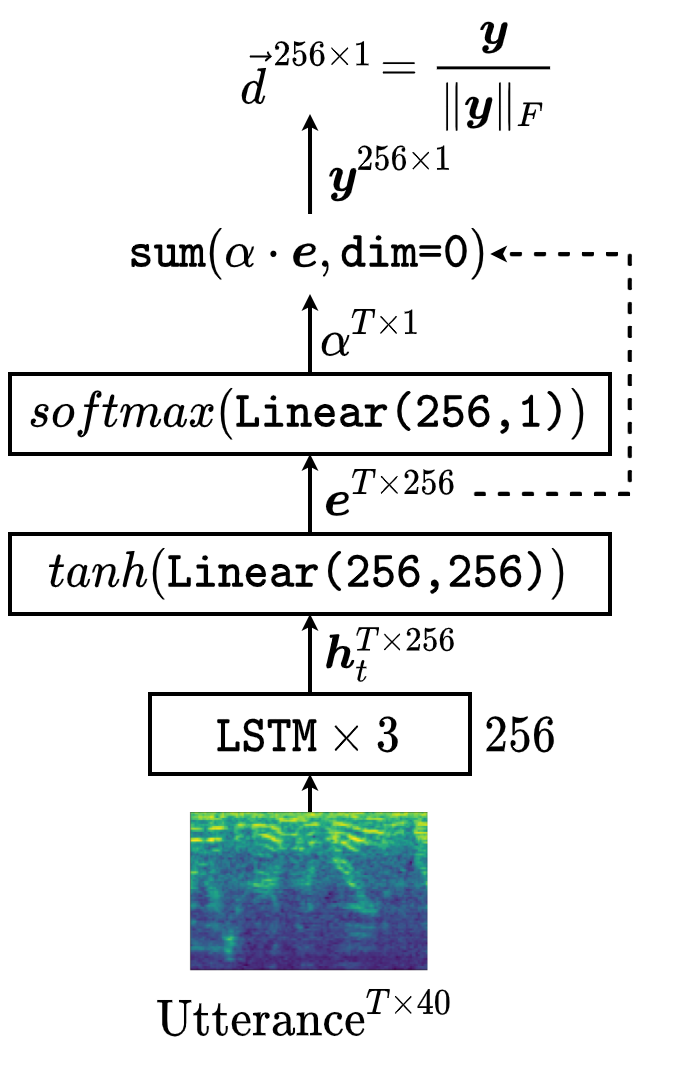}
    \caption{\textit{d}-vector}
    \label{subfig:d-vector}
  \end{subfigure}
  \hfill
  \begin{subfigure}[b]{0.67\linewidth}
    \centering
    \includegraphics[width=\linewidth]{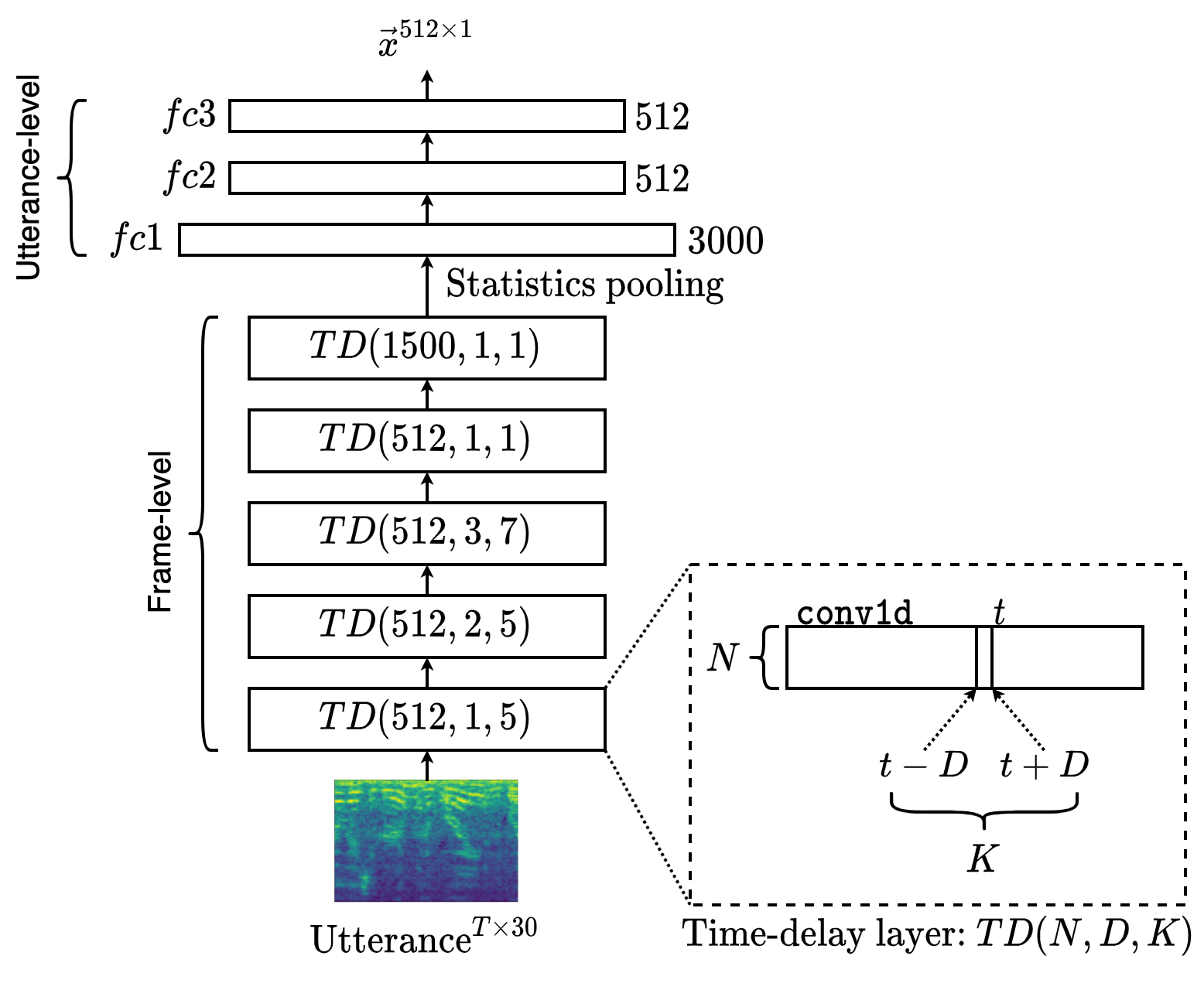}
    \caption{\textit{x}-vector}
    \label{subfig:x-vector}
  \end{subfigure}
  \caption{Vector-based speaker embedding systems}
  \label{fig:speaker_vectors}
\end{figure}

We use two additional vector-based speaker embeddings, \textit{d}-vector~\cite{variani2014deep} and \textit{x}-vector~\cite{snyder2018x}, as comparative benchmarks. Inspired by vector-based speaker verification systems such as the \textit{i}-vector~\cite{dehak2010front}, \textit{d}-vector is a DNN embedding for SV. A DNN with a 3-layer LSTM and an attentive pooling~\cite{santos2016attentive} pre-trained with the VoxCeleb1 dataset is used in our implementation. While \textit{x}-vector is also a DNN-based embedding for SV, it uses a time-delay layer~\cite{peddinti2015time} for handling short-term temporal context and a statistics pooling layer to aggregate frame-level information~\cite{snyder2017deep}. For the experiment, we use an \textit{x}-vector system pretrained for speaker recognition with VoxCeleb1 Dev and VoxCeleb 2. See Figure~\ref{fig:speaker_vectors} for the details of the \textit{d}-vector and \textit{x}-vector systems in our implementation.

For experiments, we split the dataset into the train ($N=441$) and evaluation ($N=111$) subsets while preserving the label distribution with stratification. With the training subset, we initially perform a hyperparameter search for 50 trials with the Optuna framework~\cite{akiba2019optuna} for each classifier and use the best configuration for 5-fold stratified cross-validation (CV). Finally, the predictions of 5 classifiers (each from a training fold) on the evaluation subset are combined via majority voting to obtain final predictions and evaluation scores: accuracy, f1-score, specificity, and recall. Note that micro-averaged scores are reported for all metrics except accuracy as our data is imbalanced.

As a baseline, we train four machine learning classifiers: Linear Discriminant Analysis (LDA), Support Vector Machine (SVM), Decision Tree (DT), and Random Forest (RF). We use adaptive average pooling to reshape 2-dimensional content embeddings to a 1-dimensional vector of length 512. We also train a deep linear classification head and evaluate performance once again with a 5-fold stratified CV, given the relatively small size of the dataset. We add a linear layer prior to pooling in our linear head to better incorporate the information in the time dimension of the content embedding into classification. We empirically found this results in a better classification performance than applying the pooling directly.

\section{Results and discussion}

\begin{table}[t]
\caption{Binary classification results (CN=0 vs. IM=1) with models trained w/ 5-fold stratified CV and their ensembled prediction on a test split}
\label{tab:baseline}
\centering
\begin{tabular}{@{}llllll@{}}
\toprule
 & Model & \textit{Acc.} & \textit{F1} & \textit{Spec.} & \textit{Recall} \\ \midrule
$\vec{x}^{512 \times 1}$ & LDA & 0.6306 & 0.6772 & 0.5510 & 0.6935 \\
 & SVM & 0.6036 & 0.6393 & 0.5714 & 0.6290 \\
 & DT & 0.5135 & 0.5781 & 0.4082 & 0.5968 \\
 & RF & 0.6486 & 0.7023 & 0.5306 & 0.7419 \\
 & Linear & 0.5625 & 0.5195 & 0.5327 & 0.5705 \\ \midrule
$\vec{d}^{256 \times 1}$ & LDA & 0.6396 & 0.6774 & 0.5918 & 0.6774 \\
 & SVM & 0.5675 & 0.6129 & 0.5102 & 0.6129 \\
 & DT & 0.5586 & 0.6316 & 0.4082 & 0.6774 \\
 & RF & 0.6396 & 0.6923 & 0.5306 & 0.7258 \\
 & Linear & 0.5520 & 0.4592 & 0.3474 & 0.8106 \\ \midrule
 $S^{256 \times 1}$ & LDA & 0.5946 & 0.6218 & 0.5918 & 0.5968 \\
 & SVM & 0.5676 & 0.6000 & 0.5510 & 0.5806 \\
 & DT & 0.5946 & 0.6897 & 0.3265 & 0.8065 \\
 & RF & 0.6126 & 0.6993 & 0.3673 & 0.8065 \\
 & Linear & 0.6041 & 0.5116 & 0.5814 & 0.5100 \\ \midrule
 $C^{512 \times 1}$ & LDA & 0.6396 & 0.7015 & 0.4898 & 0.7581 \\
 & SVM & 0.6036 & 0.6452 & 0.5510 & 0.6452 \\
 & DT & \textbf{0.6667} & 0.7132 & 0.5714 & 0.7419 \\
 & RF & 0.6216 & \textbf{0.7162} & 0.3265 & \textbf{0.8548} \\
 & Linear & 0.6576 & 0.6724 & \textbf{0.6938} & 0.6290 \\ \bottomrule
\end{tabular}
\end{table}

\subsection{Baseline classification}

Table~\ref{tab:baseline} shows evaluation results of models trained with 5-fold stratified CV and their ensembled prediction on a test split. Overall, content embedding models show the highest performance, while no single model outperforms others in terms of all four metrics. Interestingly, the linear classifier shows the most balanced performance across four metrics, while the RF model shows the highest F1 and recall but low specificity.

\subsection{Dependency of speaker and content embeddings}

Further, we investigate the influence of speaker embeddings on classification performance with speaker embeddings. We compare the performance of content embeddings for the CN vs. IM classification task, extracted with three different speaker encoders as the following: 1) pre-trained speaker encoder without finetuning, 2) DementiaBank finetuned speaker encoder, and 3) one-hot speaker encoding. A linear classification head was used for all three conditions for a fair comparison, with the equal learning rate of 0.004 and the AdamP optimizer.~\cite{heo2020adamp}.

\begin{table}[h]
\caption{CN vs. IM classification results using content embeddings, extracted with 3 different speaker encoders}
\label{tab:speaker-comparison}
\centering
\begin{tabular}{@{}lllll@{}}
\toprule
 & \textit{Acc.} & \textit{F1} & \textit{Spec.} & \textit{Recall} \\ \midrule
Pre-trained & 0.6250 & 0.5462 & \textbf{0.8863} & 0.4301 \\
Finetuned & \textbf{0.6576} & \textbf{0.6724} & 0.6938 & \textbf{0.6290} \\
One-hot & 0.5208 & 0.3670 & 0.8505 & 0.2594 \\ \bottomrule
\end{tabular}
\end{table}

The results in Table~\ref{tab:speaker-comparison} show that content embedding extracted with a finetuned speaker encoder achieves the best classification performance across all metrics except specificity, which is the highest with a pretrained speaker encoder. In contrast, one-hot speaker embeddings without any information except speaker identities result in the lowest performance.

Clearly, the results indicate that content embeddings' effectiveness for assessing cognitive impairments correlates with the informativeness of speaker embeddings used to extract contents. This result is unsurprising, as while the AutoVC framework claims to achieve separation of contents from the rest of information in speech with a carefully chosen bottleneck, there is a caveat as its content encoder takes both mel spectrograms and speaker embeddings as inputs, thus leaving content embeddings open to contamination considering speaker embeddings have been shown to contain content information~\cite{wang2017does,raj2019probing}. Therefore, the decoder can reconstruct speech solely from speaker embeddings if they contain enough content information, without any information encoded in content embeddings.

\subsection{Combined vs. non-combined speaker utterances}

Most speakers in the DementiaBank dataset contributed more than one speech, with multiple visits split by approximately one year. In addition, some were given different diagnoses across visits, suggesting they experienced a noticeable change in their cognition over time. Given such cases, we considered each speech sample as a unique speaker in our experiments (the augmentation setup) even though they were from the same speaker.

Nonetheless, as the information encoded in speaker embeddings was shown to affect the classification performance of content embeddings in our observation, we further test the relationship between speaker and content embeddings by combining utterances from the same speakers collected at different points in time together for the speaker encoder training and observing its influence on CN vs. IM classification with content embeddings. If the effectiveness of content embeddings for assessing cognitive impairments depends on information encoded in speaker embeddings, combining utterances of the same speakers across sessions, which will normalize the information speaker embeddings encode, should result in lower classification performance.

\begin{table}[h]
\caption{Classification results with content embeddings, combined vs. not-combined speaker utterances}
\label{tab:combined}
\centering
\begin{tabular}{@{}lllll@{}}
\toprule
 & \textit{Acc.} & \textit{F1} & \textit{Spec.} & \textit{Recall} \\ \midrule
Combined & 0.5729 & 0.5072 & \textbf{0.7767} & 0.4150 \\
Not-combined & \textbf{0.6576} & \textbf{0.6724} & 0.6938 & \textbf{0.6290} \\ \bottomrule
\end{tabular}
\end{table}

As in Table~\ref{tab:combined}, the results agree with our expectation, with content embeddings extracted under the not-combined condition resulting in a higher classification performance than the embeddings under the combined condition.

\section{Conclusions and future work}

In this paper, we used the encoder-decoder architecture based on the AutoVC framework to separately extract contents from speaker identities as embeddings and tested their viability for assessing cognitive impairments against conventional speaker embeddings using the DementiaBank Pitt Corpus. Our results show that while the content embeddings are viable for the defined problem with a comparative advantage over vector-based speaker embeddings in their representativeness with an additional time-dimension, the inherent design of the current framework for extracting content information renders them dependent on information encoded in speaker embeddings. Thus, future work may modify the current encoder-decoder architecture by removing the inflow of speaker embeddings to the content encoder and further formulate training as multi-task learning to directly assess whether a content encoder successfully extracts speech content only in the optimization process, for example, by letting a model infer content-specific components of speech such as pitch and rhythm from the embeddings~\cite{qian2020unsupervised} and include them in a loss function.

\bibliographystyle{IEEEtran}
\bibliography{mybib}

\end{document}